# J2EE™ Deployment: The JOnAS Case Study

**François Exertier**

*Bull,*
*1, rue de Provence*
*BP 208*
*38432 Echirolles Cedex*
*Francois.Exertier@objectweb.org*

RÉSUMÉ. *La spécification J2EE (Java 2 platform Enterprise Edition) définit une architecture de serveur d'application Java. Jusqu'à J2EE 1.3, seuls les aspects de déploiement concernant le développeur d'applications étaient adressés. Avec J2EE 1.4, les interfaces et les étapes de déploiement ont été plus précisément spécifiées dans la spécification "J2EE Deployment". JOnAS (Java Open Application Server) est une plate-forme J2EE développée au sein du consortium ObjectWeb. Les aspects déploiement sont en cours de développement. Cet article décrit les concepts liés au déploiement dans J2EE, ainsi que les problématiques levées lors de leur mise en œuvre pour JOnAS. Il n'a pas pour but de présenter un travail abouti, mais illustre le déploiement par un cas concret et ébauche une liste de besoins non encore satisfaits dans le domaine.*

ABSTRACT. *The J2EE (Java 2 platform Enterprise Edition) specification defines an architecture for Java Application Servers. Until J2EE 1.3, the deployment aspect was addressed from the developer point of view only. Since J2EE 1.4, deployment APIs and steps have been more precisely specified within the "J2EE Deployment Specification". JOnAS (Java Open Application Server) is a J2EE platform implementation by ObjectWeb. The deployment aspects are under development. This article describes the J2EE Deployment concepts, and the issues raised when implementing deployment features within JOnAS. It does not provide a complete solution, but illustrates deployment through a concrete example and initiates a list of non fulfilled requirements.*

MOTS-CLÉS : *J2EE, déploiement, Java, composant, serveur d'application*

KEYWORDS: *J2EE, deployment, Java, component, Application Server*

**1. Introduction**

This article describes deployment issues within a J2EE platform. After a brief introduction to J2EE and to the JOnAS implementation, the J2EE deployment features are presented, by summarizing the main concepts of the J2EE Deployment Specification (was the JSR #88 of the Java Community Process). Then the issues raised when implementing these deployment features within JOnAS are exposed, showing lacks within the standard specification, and how they may be filled by using a more complete deployment model.

**2. JOnAS and J2EE**

**2.1.** *J2EE*

The Sun$^{TM}$ J2EE specification [J2EE 03] defines an architecture and interfaces for developing and deploying distributed Internet Java$^{TM}$ server applications based on a multi-tier architecture. This specification intends to facilitate and standardize the development, deployment, and assembling of application components; such components will be deployable on J2EE platforms. The resulting applications are typically web-based, transactional, database-oriented, multi-user, secured, scalable, and portable. More precisely, this specification describes two kinds of information:

- The first is the runtime environment, called a J2EE server, which provides the execution environment and the required system services, such as the transaction service, the persistence service, the Java Message Service (JMS), and the security service.

- The second is programmer and user information explaining how an application component should be developed, deployed, and used.

Not only will an application component be independent of the platform and operating system (since it is written in Java), it will also be independent of the J2EE platform.

A typical J2EE application is composed of 1) presentation components, also called "web components" (Servlets [JSER] and JSPs$^{TM}$ [JSP]), which define the application Web interface, and 2) enterprise components, the "Enterprise JavaBeans" (EJB), which define the application business logic and application data. J2EE components may access external resources such as databases, mail servers, legacy enterprise information systems, web services, etc., through "standard" Java APIs. The J2EE server provides containers for hosting web and enterprise components. The container provides the component life-cycle management and connects the components to the services provided by the J2EE server.

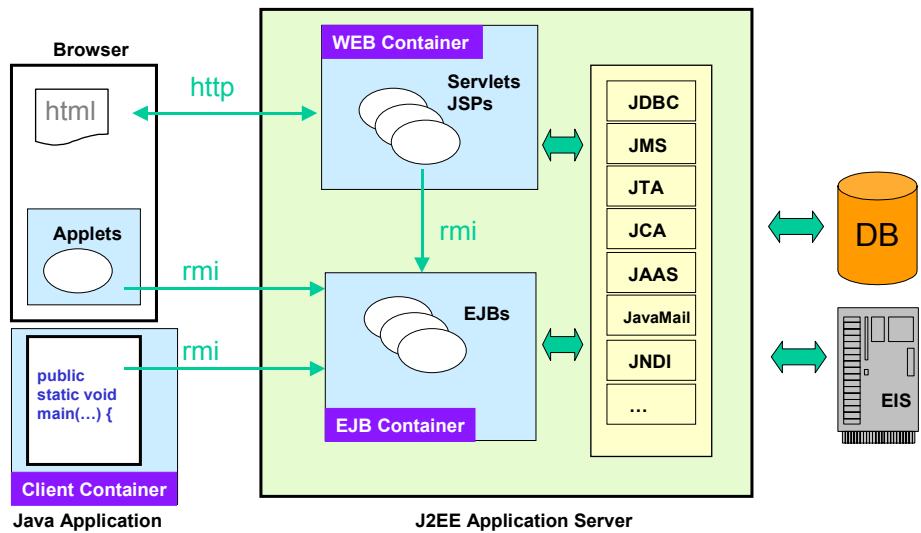

**Figure 1.** *J2EE Arhitecture*

## 2.2. *JOnAS*

JOnAS [JONA] is an open source implementation of J2EE, developed within the ObjectWeb consortium. ObjectWeb is an open source initiative which can be compared to Apache or Linux, but in the area of **middleware**. The aim of ObjectWeb is to develop and promote open source middleware software. ObjectWeb is an International Consortium hosted by INRIA, officially founded in February 2002 by Bull, France Telecom, and INRIA. All software is available with the LGPL license. The technical objective of this consortium is to develop a distributed component-based, middleware technology, in line with CORBA, Java, and W3C standards. The idea is to apply the component model, as already used at the application level in J2EE and in the CORBA Component Model, at the middleware level itself.

JOnAS is a pure Java, open source, application server currently on the way of being certified as J2EE compliant (passing the Sun Compatibility Test Suite for J2EE 1.4). JOnAS is an implementation of J2EE 1.4. It currently implements EJB 2.1. Its current integration of Tomcat or Jetty as a Web container ensures conformity to Servlet 2.4 and JSP 2.0 specifications. The JOnAS server relies on or implements the following Java APIs: J2EE CA$^{TM}$ 1.5, JDBC$^{TM}$ 3.0, JTA$^{TM}$ 1.0.1, JMS$^{TM}$ 1.1, JMX$^{TM}$ 1.2, JNDI$^{TM}$ 1.2.1, JAAS$^{TM}$ 1.0, JACC$^{TM}$ 1.0, JavaMail$^{TM}$ 1.3.

In addition to the implementation of all J2EE-related standards, JOnAS provides the following important advanced features:

**Management**: JOnAS server management uses JMX and provides a JSP/Struts-based management console.

**Services:** JOnAS's service-based architecture provides for high modularity and configurability of the server. Most of the J2EE resources are provided through JOnAS services. It also provides a way to start only the services needed by a particular application, thus saving valuable system resources. JOnAS services are manageable through JMX.

**Scalability:** JOnAS integrates several optimization mechanisms for increasing server scalability. This includes a pool of stateless session beans, a pool of message-driven beans, a pool of threads, a cache of entity beans, activation/passivation of entity beans, pools of connections (for JDBC, JMS, J2EE CA), storage access optimizations.

**Clustering:** JOnAS clustering solutions, both at the WEB (http) and EJB (rmi) levels, provide load balancing, high availability, and failover support.

**Distribution:** JOnAS works with several distributed processing environments, due to the integration of the CAROL [CARO] (Common Architecture for RMI ObjectWeb Layer) ObjectWeb project, which allows simultaneous support of several communication protocols: RMI/JRMP the Sun proprietary protocol, RMI/IIOP, CMI the "Cluster aware" distribution protocol of JOnAS, Jeremie an optimized implementation of RMI by ObjectWeb. Used with Jeremie or JRMP, JOnAS benefits from transparent local RMI call optimization.

**Support of "Web Services:"** Due to the integration of AXIS, JOnAS allows J2EE components to access "Web services" (i.e., to be "Web Services" clients), and allows J2EE components to be deployed as "Web Services" endpoints. Standard Web Services clients and endpoints deployment, as specified in J2EE 1.4, is supported.

**Support of JDO:** By integrating the ObjectWeb implementation of JDO [JDO], SPEEDO [SPDO], and its associated J2EE CA Resource Adapter, JOnAS provides the capability of using JDO within J2EE components.

## 3. J2EE Deployment

The J2EE Deployment Specification [JDEP 03], available since J2EE 1.4, allows a clear separation between the J2EE platform and the tool used to deploy applications upon such a platform. The specification defines APIs to be provided/required by deployment tools and J2EE products, in order for any compliant deployment tool to be able to deploy applications on any compliant J2EE product.

Among all the specified interfaces, the most important one to understand the principle of this architecture, is the DeploymentManager, which is to be provided by the J2EE product, and around which the interaction between the deployment tool and the J2EE platform is performed. The DeploymentManager deals with EAR, JAR, WAR and RAR files, which are the J2EE standard deployable units:

- EJB-JAR files are Java archives containing EJBs classes and xml deployment descriptors

- WAR files are Java archives containing Web applications, i.e. Servlets/JSP, libs, and xml deployment descriptors)

- EAR files are Java archives containing EJB-JARs, WARs, libs, and xml deployment descriptors; they generally represent J2EE applications

- RAR files are Java archives containing J2EE CA Connector classes and xml deployment descriptors; connectors are software adapters for accessing the Enterprise Information System from J2EE component through Java APIs, security, connection and transaction stuff being transparently managed by the application server.

The Deployment Manager provides interfaces for

- Configuring an application: generating the deployment descriptors info needed by the application server. This first step is also called "deployment", and consists in configuring a "unit" (ear, war, ejb-jar, rar). This is the job of the deployer or application assembler, in charge of configuring the "units" for their specific use within a given instance of the application (e.g. establishing links between components), and for the target platform (e.g. specify a particular resource/service to be used on the platform). This is done by updating the xml deployment descriptor, and should be performed by using a "deployment tool" (generally graphical) relying on the J2EE Deployment API for interacting with a given J2EE platform; the tool asks the platform for the server specific required deployment information, and interacts with the deployer to fill the corresponding elements. The generation of the J2EE platform container specific interposition classes (stubs) is generally part of this step. The result is a deployed unit, an archive file which contains what was in the "input" unit, plus the interposition classes.

- Distributing an application over a target (see below) consists in installing "deployed" units (i.e. configured units) on a given target. A target is one or several J2EE servers

- Starting/Stopping an application

- Undeploying an application
- Monitoring/Managing deployed modules.

Operations of the DeploymentManager for distributing, starting/stopping, undeploying an application, apply on a *target*, which may be either a single application server, or a group of application servers. A TargetModuleID object is a reference to a J2EE module that has been deployed to a Target.

For operations such as distribute, start/stop, undeploy, the DeployerManager returns a ProgressObject to the management tool, which is an object for tracking and reporting the execution of potentially long-lived deployment activities. The ProgressObject is either polled by the tool (e.g. calling getDeploymentStatus, …), or it may raise events toward the tool (which registers a callback handler: ProgressListener).

A DeploymentManager may be used in disconnected mode, in this case only for configuring modules.

The figure below shows a subset of the different APIs (the main ones) that should be provided by the tool provider and the J2EE product provider.

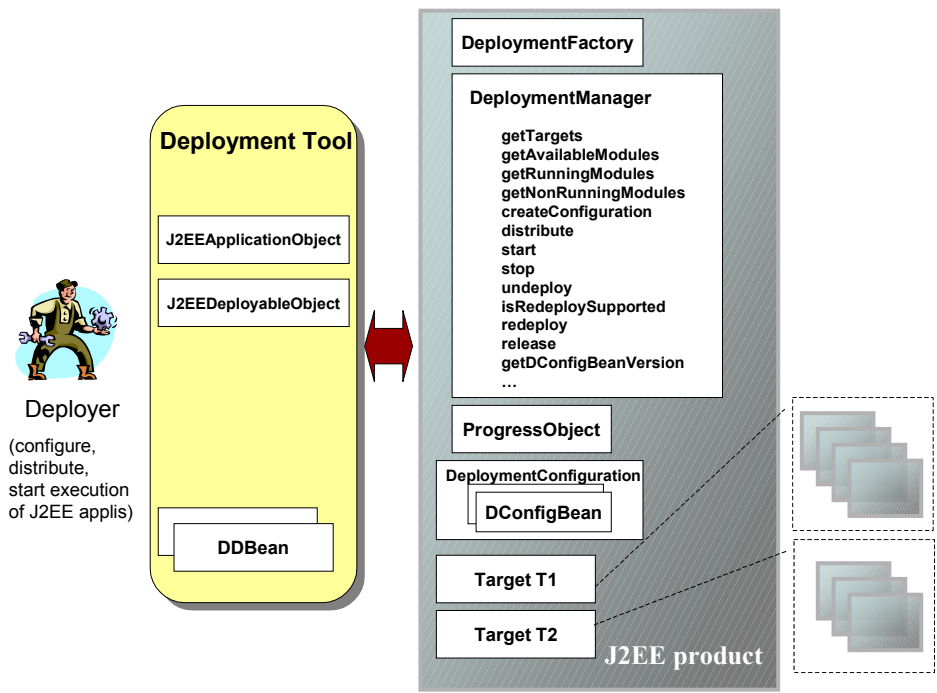

**Figure 2**. *J2EE Deployment*

In order to understand the use of these APIs, an example of interaction scenario between the deployment tool and the J2EE platform is following:

- The Deployment tool gets a J2EE module to be deployed: if the module is a WAR, a JAR or a RAR, it instanciates a *J2EEDeployableObject*, if the module is an EAR it instanciates a *J2EEApplicationObject* (containing *J2EEDeployableObject*s)

- The Deployment tool makes use of the *DeploymentFactory* to get a *DeploymentManager* object

- The Deployment tool calls the *createConfiguration* method of the *DeploymentManager* with the *J2EEDeployableObject* as parameter

- The *DeploymentManager* returns the *DeploymentConfiguration* object containing the *DConfigBean* objects, representing the platform specific deployment descriptors info (e.g. the jonas-ejb.xml representation)

- The Deployment tool lets the deployer fill all necessary deployment elements (platform specific -> *DConfigBean*, and standard ->*DDBean*). In fact the *DDbean* (resp. *DConfigBean*) objects are JavaBeans used for introspecting standard (resp. platform specific) deployment descriptors and for presenting to the deployer a logical grouping of deployment (resp. configuration) information. A J2EE module's deployment descriptor is represented by a composition of *DDBean*s (or *DConfigBean*s).

- Then the Deployment tool (i.e. the deployer through the tool) calls the *distribute* method of the *DeploymentManager* object in order to make the application available on the servers specified by a *target* object. A *TargetModuleId* is associated to the deployment of this module on the specific target.

## 4. Deployment Issues

### 4.1. *JOnAS implementation*

Implementing the J2EE Deployment Specification for JOnAS consists in implementing the product provider interfaces, i.e. the DeploymentManager and all related interfaces, which is tightly linked to the JOnAS implementation. These interfaces implementations may be compared to an adapter between the deployment tool relying on standard deployment interfaces, and the application server specific deployment mechanisms. For example, the deployment step will make use of the JOnAS GenIC tool, to generate interposition classes (from the J2EE component classes and their associated fulfilled deployment descriptors), in order to obtain the "deployed applications". The "distribute" operation will make use of the JOnAS management tool for installing a "deployed application" on a running JOnAS server. JOnAS implements the J2EE Management Specification [JMGT 03] (was the JSR #77 of the Java Community Process): the start/stop operations and the monitoring/managing of deployed modules operations will thus rely on the State Management features and on the J2EEDeployedObject managed object defined in this specification. For example it may use J2EEDeployedObject to get the list of deployed modules. These operations will be available through the Management EJB (MEJB).

This Deployment APIs implementation will be provided as a "driver" (a jar file), that may be plugged to any standard deployment tool intending to deploy applications on JOnAS.

**4.2.** *Issues and Related Work*

The J2EE Deployment Model however does not provide solutions to all the deployment problems. This section lists some of the issues that were raised when experimenting the J2EE Deployment Model on JOnAS.

The J2EE Deployment model does not provide the capability to express dependencies between deployed modules. Generally a J2EE application packaged within an EAR file is self-contained, i.e. it will not use components of other packages. However the possibility is left open by J2EE, i.e. it is possible to deploy several modules (ear, war, ejb-jar, rar), that may use each other ... For example a connector (RAR) may be deployed for being used (and shared) by several applications, in this case it is possible to deploy and install it standalone, i.e. not packaged within an EAR file. However nothing prevents an EAR to be installed if a required module such an RAR has not been previously installed.

More and more, J2EE resources may be deployed as standard deployment units: JDBC datasources, JMS Connection Factories, may be deployed as J2EE CA connectors (RAR units). However there are still some resources that are made available through ad hoc means, mainly server configuration steps: this is the case for Mail factories, which are provided through server configuration. For example, a J2EE application may need that the JOnAS Mail Service be launched and that a given Mail Factory be created. The J2EE Deployment Model does not allow to specify such dependencies, neither to express some requirements on the target server configuration. Dependencies on J2EE services could be used in many cases with JOnAS, which currently does not provide dynamic service launching facilities (launch a service when needed). Dependencies between applications and JOnAS services to be launched may be expressed for some of the JOnAS services: Mail, EJB container, Web container, and Web Services services. This is currently no more necessary for other services: JMS and Database JOnAS services will be deprecated (replaced by the corresponding connectors which are deployable units), while Transaction, Communication (registry), Security, and Ear services are or will become mandatory services.

The J2EE Deployment model defines a simple concept of target, which may be a server, or a set of servers. This may be enough to express deployment/installation operations on top of a cluster of J2EE servers. However there are some complex distributed configurations that may not be directly processed by such a deployment scheme. E.g. when a component running on a site A should access a component running on a remote site B, it should be possible to specify that the JOnAS communication service of the server running on site A should use the registry (communication service) of the server running on site B.

All these issues may be solved by using a more generic deployment model, such as the one provided through the Generic Deployment Framework [DON 04]. This model defines deployment units, applications, sites, and dependencies. The dependency concept is generic enough to express any kind of dependency, between deployment units, site (i.e. server) configuration, … It is also to be noted that such a deployment framework is not restricted to a given type of platform, it should be possible to use it for heterogeneous deployment, e.g. for an application with parts running on J2EE, CORBA and OSGi. A special attention should also be delivered to the OMG specification for Deployment and Configuration (D&C) of Component-based Distributed Applications [OMG 03], which is in the state of an adopted draft. The OMG D&C specification defines a platform independent model to address most of the deployment issues. A mapping of this model for the CORBA CCM platform is provided in this specification, a mapping for J2EE may certainly help solving most of the issues raised in the JOnAS case.

## 5. Conclusion

Deployment on application servers is an important topic that is handled in a specific way for each kind of platform. The J2EE deployment concepts and process have been precisely defined within the J2EE Deployment Specification. After a short introduction to J2EE and to the JOnAS platform, this paper presents the J2EE deployment concepts, their implementation within JOnAS, and the restrictions of the corresponding model. Although the J2EE deployment concepts are well specified and allow a clear separation between J2EE platforms and deployment tools, this is not enough for covering all the deployment concerns. The aim of this paper is clearly to illustrate application server deployment through the J2EE example and to shortly demonstrate the utility of a more generic deployment model, not necessarily restricted to J2EE.

## 6. Bibliographie

[CARO]   *Common Architecture for RMI ObjectWeb Layer (CAROL)*. http://carol.objectweb.org.

[DON 04] Didier Donsez, Vincent Lestideau, *Osmose/ObjectWeb Generic Deployment Framework Specification, v 0.1*, UJF-IMAG-LSR-ADELE, June 2004, available at http://www-adele.imag.fr/~donsez/ow/owdeploy/doc/OSMOSE-ObjectWeb-Generic_Deployment_Framework.html.

[J2EE 03] *Java™ 2 Platform, Enterprise Edition Specification Version 1.4*. November 2003. Available at http://java.sun.com/j2ee/docs.html.

[JCA 03] *J2EE™ Connector Architecture 1.5* (Connector specification). November 2003. Available at http://java.sun.com/j2ee/connector.

[JDEP 03] *Java™ 2 Platform, Enterprise Edition Deployment Specification 1.0* (J2EE Deployment specification). November 2003. Available at http://jcp.org/jsr/detail/88.jsp.

[JDO] *Java Data Objects (JDO) API and Specification*. http://java.sun.com/products/jdo.

[JMGT 02] *Java™ 2 Platform, Enterprise Edition Management Specification 1.0* (J2EE Management specification). July 2002. Available at http://jcp.org/jsr/detail/77.jsp.

[JONA] ***J****ava **O**pen **A**pplication **S**erver (JOnAS)*. http://jonas.objectweb.org.

[JSER] *Java™ Servlet Technology*. http://java.sun.com/products/servlet.

[JSP] *Java™ ServerPages (JSP) Technology*. http://java.sun.com/products/jsp.

[OMG 03] *Deployment and Configuration of Component-based Distributed Applications Specification*. OMG Adopted Draft. June 2003. Available at http://www.omg.org/docs/ptc/03-07-02.pdf.

[SPDO] *ObjectWeb implementation of JDO (Speedo)*. http://speedo.objectweb.org.